\documentclass[dvips]{imsart}

\usepackage{amsmath,amssymb,float,natbib,graphicx}
\usepackage[thmmarks,noconfig]{ntheorem}
\usepackage[scale=0.7]{geometry}
\theoremnumbering{arabic}
\theoremstyle{plain}
\RequirePackage{latexsym}
\theoremsymbol{} 
\theorembodyfont{\itshape}
\theoremheaderfont{\normalfont\bfseries}
\theoremseparator{}

\newtheorem{theorem}{Theorem}

\newtheorem{proposition}{Proposition}

\newcommand{\tr}{\mathrm{trace}}
\newcommand{\V}{\mathrm{var}}
\newcommand{\E}{\mathbb{E}} 
\newcommand{\R}{\mathbb{R}}
\newcommand{\HH}{\mathcal{H}}

\begin{document}
\begin{frontmatter}
\title{Recursive Bias Estimation for multivariate regression smoothers}
\runtitle{Recursive Bias Estimation}

\begin{aug}
\author{\fnms{Pierre-Andr\'e} \snm{Cornillon}\ead[label=e1]{pierre-andre.cornillon@supagro.inra.fr}},
\author{\fnms{Nicolas} \snm{Hengartner}\ead[label=e2]{nickh@lanl.gov}}
\and
\author{\fnms{Eric} \snm{Matzner-L\o ber}\corref{}\ead[label=e3]{eml@uhb.fr}}
\affiliation{Montpellier SupAgro, University Rennes 2 and Los Alamos National Laboratory}

\address{Address of P-A Cornillon\\
Statistics, IRMAR UMR 6625,\\
Univ. Rennes 2, \\
35043 Rennes, France\\
\printead{e1}}

\address{Address of N. Hengartner\\
Los Alamos National Laboratory,\\
NW, USA\\
\printead{e2}
}

\address{Address of E. Matzner-L\o ber\\
Univ. Rennes, \\
35043 Rennes, France\\
\printead{e3}\\
}

\end{aug}

\begin{abstract}%
This paper presents a practical and simple fully nonparametric 
multivariate smoothing procedure that adapts to the underlying smoothness 
of the true regression function.  Our estimator is easily computed
by successive application of existing base smoothers (without the need of
selecting an optimal smoothing parameter), such as 
thin-plate spline or kernel smoothers.  
The resulting smoother has
better out of sample predictive capabilities than the underlying 
base smoother,
or competing structurally constrained models (GAM) for small
dimension ($3 \leq d \leq 7$) and moderate sample size $n \leq 800$.
Moreover our estimator is still useful when $d> 10$ and to our knowledge,
no other adaptive fully nonparametric  regression 
estimator is available without constrained
assumption such as additivity for example.  On a real example, the 
Boston Housing Data, our method reduces the out of sample prediction 
error by 20\%.  An R package {\bf ibr},
available at CRAN, implements the proposed multivariate nonparametric method 
in R.
\end{abstract}

\begin{keyword}[class=AMS]
\kwd{62G08}
\end{keyword}

\begin{keyword}
\kwd{nonparametric regression}
\kwd{smoother}
\kwd{kernel}
\kwd{thin-plate splines}
\kwd{stopping rules}
\end{keyword}
\end{frontmatter}

\section{Introduction}
Regression is a fundamental data analysis tool for uncovering functional 
relationships between pairs of observations $(X_i,Y_i), i=1,\ldots,n$. The 
traditional approach specifies a parametric family of regression functions to
describe the conditional expectation of the response variable $Y$ given the 
independent multivariate variables $X \in \R^d$, and estimates 
the free parameters 
by minimizing the squared error between the predicted values and the data.
An alternative approach is to assume that the regression function varies
smoothly in the independent variable $x$ and then estimate locally the conditional 
expectation $m(x)={\mathbb E}[Y|X=x]$. This results in nonparametric regression 
estimators.  We refer the interested reader to
\citet{eubank1999}, \citet{fan+1996} and \citet{simonoff1996}
for a more in depth treatment of various classical regression smoothers.  
The vector of predicted
values $\widehat Y_i$ at the observed covariates $X_i$ from a nonparametric 
regression is called a regression smoother, or simply a smoother, because the 
predicted values $\widehat Y_i$ are less variable than the original observations
$Y_i$.   
Operationally, linear smoothers can be written as
\begin{eqnarray*}
\widehat m = S Y,
\end{eqnarray*}
where $S$ is a $n \times n$ smoothing matrix.  Smoothing matrices $S$
(or $S_\lambda$) typically depend on a tuning parameter, which we
denote by $\lambda$, that governs the tradeoff between the smoothness
of the estimate and the goodness-of-fit of the smoother to the data,
by controlling the effective size of the local neighborhood of the
explanatory variable over which the responses are averaged.  We
parameterize the smoothing matrix such that large values of $\lambda$
will produce very smooth curves while small $\lambda$ will produce a
more wiggly curve that wants to interpolate the data.  For example,
the tuning parameter $\lambda$ is the bandwidth for kernel smoother,
the span size for running-mean smoother, and the scalar that governs
the smoothness penalty term for thin plate splines (TPS).

It is well known that given $n$ uniformly distributed points in the unit
cube $[-1,1]^d$, the expected number of points that are covered 
by a ball centered at the origin with radius $\varepsilon < 1$, scales
as $n \varepsilon^d$.  This is to say that covariates in high dimensions
are typically sparse.   This phenomenon is sometimes called \textit{the
curse of dimensionality}.   As a consequence, nonparametric smoothers
must average over larger neighborhoods, which in turn produces
more heavily biased smoothers.   Optimally selecting the
smoothing parameter does not alleviate this problem.  Indeed, 
when the regression function $m$ mapping ${\mathbb R}^d$ to ${\mathbb R}$
belongs to some finite smoothness functional classes (H\"older, Sobolev, Besov)
the optimal mean squared error rate of convergence is 
$n^{-2 \nu/(2 \nu +d)}$ where $\nu$ is the smoothing index \citep[see for
example][]{tsybakov2009}.  
Common wisdom suggest avoiding general nonparametric smoothing in 
moderate dimensions (say $d > 5$) and focus instead on fitting structurally 
constrained regression models, such as additive 
\citep{hastie+1995,linton+1995,hengartner+2005} and projection pursuit
models \citep{friedman+1981}.   The popularity of additive models stems 
in part from the interpretability of the individual estimated additive 
components, and from the fact that the estimated regression function converges
to the best additive approximation of the true regression function at the optimal
univariate mean squared error rate of $n^{-2 \nu/(2 \nu +1)}$.
While additive models do not estimate the true underlying regression function, 
one hopes for the approximation error to be small enough so that 
for moderate sample sizes, the prediction mean square error 
of the additive model is less than the prediction error of a fully nonparametric 
regression model. 

The impact of the curse of dimensionality is lessened for very smooth 
regression functions. For regression functions with $\nu=2d$ continuous derivatives, 
the optimal rate is  $n^{-4/5}$, a value recognized as the optimal mean squared error of 
estimates for twice differentiable univariate regression functions.
The difficulty is that in practice, the smoothness of the regression function
is typically unknown.  Nevertheless, there are large potential gains
(in terms of rates of convergence) if one considers multivariate smoothers
that adapt to the smoothness of the regression function. 
Since the pioneer work of \citet{lepski1991},
adaptive nonparametric estimation became a major topic in mathematical
statistics (see for example \citealt{gyorfi+++2020} or \citealt{tsybakov2009}).
Adaptive nonparametric estimator can be achieve either by direct estimation 
(see Lepski's method and related papers) or by aggregation of different 
procedures \citep[see][]{yang2000}. This paper presents a practical and simple nonparametric 
multivariate smoothing procedure that adapts to the underlying smoothness 
of the true regression function.  Our estimator is easily computed
by successive application of existing smoothers, such as Thin Plate Spline  
or kernel smoother.  Thanks to adaptivity (proven for TPS smoother), our estimator behaves 
nicely in small dimension ($3 \leq d \leq 7$) with moderate sample size $n \leq 800$
and remains useful when $d> 10$.

Section 2 introduces our procedure and motivates it as repeated
corrections to the bias of a smoother, where at each step, the bias
is estimated by smoothing the residuals.  We use the Generalized 
Cross-Validation (GCV) criteria to stop our iterative procedure when the
prediction error of our estimate is nearly minimized.
The idea of estimating the bias from residuals to correct a pilot
estimator of a regression function goes back to the concept of
\textit{twicing} introduced by \citet{tukey1977} to estimate bias
of misspecified multivariate regression models.  Numerous authors have
shown the benefits of various bias reduction techniques in 
nonparametric regression, including \citet{he+2009,choi++2000,
choi+1998,hengartner+++2010,hirukawa2010}.

The idea of iterative debiasing regression smoothers is
already present in \citet{breiman1999} in the context of the \textit{bagging}
algorithm.  More recently, the interpretation of 
the $L_2$-boosting algorithm as an iterative bias correction scheme  
was alluded to in \citet{ridgeway2000}'s discussion of 
\citet{friedman++2000} paper on the statistical interpretation of boosting.
\citet{buhlmann+2003} present the statistical properties of the $L_2$-boosted 
univariate smoothing splines and proposed an additive procedure to deal with 
multivariate data. \citet{marzio+2008} describes the behavior of univariate 
kernel smoothers after a single bias-correction iteration.

Section 3 applies the iterative bias reduction procedure 
to multivariate Thin Plate Spline smoothers.  TPS smoothers have
attractive theoretical properties that facilitate our proofs of 
the adaptation to the unknown smoothness of our procedure.  However,
implementation of the TPS is limited by the need of the sample size to be 
larger than size of its parametric component.  The latter grows
exponentially with the dimension of the covariates $d$. For practical 
considerations, we consider, in Section 4, the iterative bias reduction 
procedure using kernel smoothers that can be applied more generally than TPS.  
We discuss the use of different kernels since the choice of the kernel
is important for the iterative bias reduction procedure. 

The simulation results presented in Section 5
show that for moderate dimensions of the covariates (eg. $3 \leq d \leq 7$), and 
sample sizes ranging from $n=50$ to $n=800$, our iterated smoother has 
significantly smaller prediction error than the base smoother with using 
an ``optimal smoothing'' parameter.  We end this section with the prediction 
of the classical Boston housing data set ($n=506$ and $d=13$).
The interested reader can download an R implementation of our procedure 
with optimized computations for moderate sample size \citep{cornillon++2010a}. 

Finally, the proofs are gathered in the Appendix.

\section{Iterative bias reduction} \label{section:bias}

This section presents the general iterative bias reduction framework
for linear regression smoothers and shows that the resulting smoother,
when combined with GCV, adapts to the underlying smoothness of the 
regression function.  The advantage of our smoother is its simplicity:
we only need to repeatedly smooth residuals using existing multivariate
smoothers.  The cost of adaptation is an increase in computational 
complexity.  

\subsection{Preliminaries}
Suppose that the pairs  
$(X_i,Y_i) \in {\mathbb R}^d \times {\mathbb R}$ are related through the 
regression model
\begin{eqnarray} \label{eq:basic.model}
Y_i &=& m(X_i) + \varepsilon_i, \quad i=1,\ldots,n,
\end{eqnarray}
where $m(\cdot)$ is an unknown smooth function, and the disturbances $\varepsilon_i$ 
are independent mean zero and variance $\sigma^2$ random variables that are 
independent of all the covariates $(X_1,\ldots,X_n)$.
It is helpful 
to rewrite Equation (\ref{eq:basic.model})  in vector form by 
setting $Y=(Y_1,\ldots,Y_n)^\prime$, $m=(m(X_1),\ldots,m(X_n))^\prime$ and 
$\varepsilon=(\varepsilon_1,\ldots,\varepsilon_n)^\prime$, to get 
\begin{eqnarray}
Y &=& m + \varepsilon.   \label{eq:model.vector}
\end{eqnarray}
Linear smoothers can be written as
\begin{eqnarray} \label{eq:smoother.0}
\widehat m = S_\lambda Y,
\end{eqnarray}
where $S_\lambda$ is an $n \times n$ smoothing matrix and 
$\widehat m  =(\widehat Y_1,\ldots,\widehat Y_n)^\prime$, 
denotes the vector of fitted values. From now on, we denote 
the smoothing matrix by $S$. Let $I$ be the $n \times n$ identity matrix.  
The bias of the linear smoother (\ref{eq:smoother.0}), 
conditionally on the observed values of the covariates 
$X_1^n = (X_1,\ldots,X_n)$,  is
\begin{eqnarray}
\E[\widehat m|X_1^n] - m &=& (S-I)m= -\E[(I-S)Y|X_1^n]. \label{eq:bias.0b}
\end{eqnarray}

\subsection{Bias reduction of linear smoothers}
Expression (\ref{eq:bias.0b}) for the bias suggests that it can be estimated
by smoothing the negative residuals $-R=-(Y-\widehat m)=-(I-S)Y$. An alternative 
approach is to estimate the bias by plugging in an estimator for the regression 
function $m$ into the expression (\ref{eq:bias.0b}). The resulting estimators
are different except if we consider using the same smoother for estimating the bias
and for estimating the initial smoother. From now on, we consider using the same 
smoother. The initial estimator is given by
\begin{eqnarray*}
\hat m_1 &=& S Y := S_1 Y.
\end{eqnarray*}
Smoothing the residuals
\begin{eqnarray*} 
\hat b_1 := -S R_1 = -S(I-S_1)Y
\end{eqnarray*}
estimates the bias. Correcting the initial smoother $\widehat m_1$ by subtracting 
$\hat b_1$ yields a \textit{bias corrected} smoother
\begin{eqnarray*}
\widehat m_2 &=& S_1Y - \hat b_1\\
&=& S_1Y + S(I-S_1)Y := S_2 Y.
\end{eqnarray*}
Since $\hat m_2$ is itself a linear smoother, it is possible 
to correct its bias as well.  Repeating the bias reduction step $k-1$
times produces  the linear smoother at iteration $k$:
\begin{eqnarray*}
\widehat m_k &=& S_{k-1}Y + S(I-S_{k-1})Y :=S_{k-1}Y - \hat b_{k-1}:= S_k Y
\end{eqnarray*}
The resulting $k^{th}$ iterated bias corrected smoother becomes
\begin{eqnarray} \label{eq:mk}
\hat m_k =  [I - (I-S)^k]Y := S_k Y.
\end{eqnarray}
In the univariate case, smoothers of the form
(\ref{eq:mk}) arise from the $L_2$-boosting algorithm when 
setting the convergence factor $\mu_k$ of that algorithm to one.
Thus we can interpret the $L_2$-boosting algorithm as an
iterative bias reduction procedure.  
From that interpretation,
it follows that the $L_2$-boosting of  projection smoothers, as is the 
case for polynomial regression, bin smoothers and regression splines, is 
ineffective since the estimated bias
\[
\hat b = S(I-S)Y = 0.
\]

\subsection{Predictive smoothers}
Our smoothers predict the conditional expectation of responses only at the design points.  
It is useful to extend regression smoothers to enable 
predictions at arbitrary locations $x \in {\mathbb R}^d$ of the covariates.  
Such an extension allows 
us to assess and compare the quality of various smoothers
by how well the smoother predicts new observations.
To this end, write the prediction of the linear smoother $S$
at an arbitrary location $x$ as
\[
\hat m(x) = S(x)^\prime Y,
\]
where $S(x)$ is a vector of size $n$ whose entries are the weights
for predicting $m(x)$.  The vector $S(x)$ is readily computed for 
many of the smoothers used in practice.
Next, writing the iterative bias corrected smoother $\widehat m_k$
as
\begin{eqnarray*}
\widehat m_k &=& \widehat m_1 -\widehat b_1 + \dots - \widehat b_{k-1} \\
&=& S[I+(I-S)+(I-S)^2+\dots+(I-S)^{k-1}]Y\\
&=& S \widehat \beta_k,
\end{eqnarray*}
it follows that we can predict $m(x)$ by
\begin{eqnarray}
\label{entoutx}
\widehat m_k(x) = S(x)^\prime \widehat \beta_k.
\end{eqnarray}

\subsection{Properties of iterative bias corrected smoothers}
The squared bias and variance of the $k^{th}$ iterated bias corrected 
smoother $\widehat m_k$ (\ref{eq:mk}) are
\begin{eqnarray*}
\left(\E [\hat m_k|X_1^n] - m\right)^2 &=& m^\prime\left((I-S)^k\right)^\prime(I-S)^k m\\
\V(\hat m_k|X_1^n) &=& \sigma^2(I-(I-S)^k)\left((I-(I-S)^k)\right)^\prime,
\end{eqnarray*}
This shows that the qualitative behavior of the sequence of iterative bias corrected 
smoothers $\widehat m_k$ can be related to the spectrum of $I-S$.  The next proposition 
collects the various results for sequence of iterated bias corrected linear 
smoothers.

\begin{proposition} \label{theorem:converge}  Suppose that the singular values 
$\lambda_j$ of $I-S$ satisfy
\begin{eqnarray} \label{eq:condition.theorem}
0 \leq \lambda_j \leq 1 \quad \mbox{for} \quad j=1,\ldots,n.
\end{eqnarray}
Then we have that
\begin{eqnarray*}
&& \|\hat b_k\| < \|\hat b_{k-1}\| \quad \mbox{ and } \quad \lim_{k \rightarrow \infty} 
\hat b_k = 0,\\
&& \lim_{k \rightarrow \infty} \widehat m_k = Y \quad \mbox{ and } \quad 
\lim_{k \rightarrow \infty} {\mathbb E}[\|\widehat m_k - m\|^2|X_1^n] = n \sigma^2.
\end{eqnarray*}
\end{proposition}
The assumption that for all $j$, the singular values $0 \leq \lambda_j \leq 1$ implies that 
$I-S$ is a contraction, so that $\|(I-S)Y\| < \|Y\|$.  This condition however does not 
imply that the smoother $S$ is itself a shrinkage smoother as defined by \citet{buja++1989}.
Conversely, not all shrinkage smoothers satisfy condition (\ref{eq:condition.theorem}) of 
the theorem.  In Sections 3 and 4, we give examples of common shrinkage smoothers for which 
$\lambda_j > 1$, and show numerically that for these shrinkage smoothers, the iterative 
bias correction scheme fails.  

The proposition indicates that the number of iterations of the bias correction scheme 
is analogous to smoothing parameters of more classical smoothers:  For small numbers
of iterations, the smoother is very smooth, becoming increasingly wiggly as the number of
iterations increases, to ultimately interpolate the data.  Smoothers at either 
extreme (oversmoothing or interpolating the data) may have large prediction errors, 
and the presumptions is that along the sequence of bias corrected smoother smoothers, 
there will be smoothers that have significantly smaller prediction errors. 
In Section 3, we show no only that this fact holds for thin plate smoothing 
splines, but that there exists smoothers in that sequence that "adapts to the 
unknown smoothness" of the regression function and achieves the optimal rate of 
convergence. Since standard thin plate spline smoothers are not adaptive, 
this demonstrates the usefulness of iterative bias correction. 

\subsection{Data-driven selection of the number of steps}

The choice of the number of iterations is crucial since each iteration of the 
bias correction algorithm reduces the bias and increases the variance. 
Often a few iterations of the bias correction scheme will improve upon  the 
pilot smoother. This brings up the important question of how to decide when
to stop the iterative bias correction process.   

Viewing the latter question as a model selection problem suggests  stopping rules 
for the number of iterations based
on Akaike Information Criteria (AIC) \citep{akaike1973}, modified AIC 
\citep{hurvich++1998}, Bayesian Information Criterion (BIC) \citep{schwarz1978}, 
cross-validation, L-fold cross-validation,  Generalized cross validation 
\citep{craven+1979}, and data splitting \citep{hengartner++2002}. Each of these 
data-driven model selection methods estimate an 
optimum number of iterations $k$ of the iterative bias correction algorithm 
by minimizing estimates for the expected squared prediction 
error of the smoothers over some pre-specified set ${\mathcal K_n}=\{1,2,\ldots,M_n\}$
for the number of iterations.  

Extensive simulations of the above mentioned model selection criteria, both in 
the univariate and the multivariate settings \citep{hengartner++2008b} have shown 
that GCV 
\begin{eqnarray*}
\hat k_{GCV}&=&\arg\min_{k \in \mathcal{K}}\left\{ \log{\widehat
{\sigma_k}^2}-2\log{\left(1-\frac{\tr(S_k)}{n}\right)} \right\}
\end{eqnarray*}
is a good choice, both in terms of computational efficiencies and 
of producing good  final smoothers and asymptotic results (cf Theorem
\ref{li}). At each iteration, $\widehat{\sigma_k}^2$ corresponds 
to the estimated variance of the current residuals.

Strongly related to the number of iteration is the smoothness of the pilot
smoother, since the smoother the pilot is, the bigger is the number of 
iteration. This point and the algorithm used to select the number of 
iteration are not developed in this paper but are presented in greater detail
in the companion paper related to the R-package. However, one has to be
sure that the pilot smoother oversmooths. We will discuss that point in
the simulation part, since it depends on the type of smoother (thin plate
spline, kernel).

\section{Iterative bias reduction of multivariate thin-plate splines smoothers}
\label{section:spline}

We study the statistical properties of the
iterative bias reduction of multivariate thin-plate spline smoothers.
Given a smoothing parameter $\lambda$, the thin-plate smoother of 
degree $\nu_0$ minimizes
\begin{eqnarray}
\label{eq:def.spline}
\min_{f}
\sum_{i=1}^n \left ( Y_i - f(X_i) \right )^2 +
\lambda \left[
\sum_{\begin{tiny} \begin{array}{c} i_1,\ldots,i_d \geq 0  \\ i_1+\dots+i_d \leq \nu_0 \end{array} \end{tiny}} \int_{\R^d} 
\left|\frac{\partial^{i_1+\dots+i_d} }{\partial x_{i_1}\ldots \partial x_{i_{\nu_0}}} f(x) \right|^2 dx
\right].
\end{eqnarray}
Thin-plate smoothing splines are an attractive class of multivariate 
smoothers for two reasons: first, the solution of (\ref{eq:def.spline}) 
is numerically tractable \citep[see][]{gu2002}, and second, the eigenvalues 
of the smoothing matrix are approximatively known \citep[see][]{utreras1988}.   

\subsection{Numerical example}
\begin{figure}[H] 
\begin{center}
\includegraphics{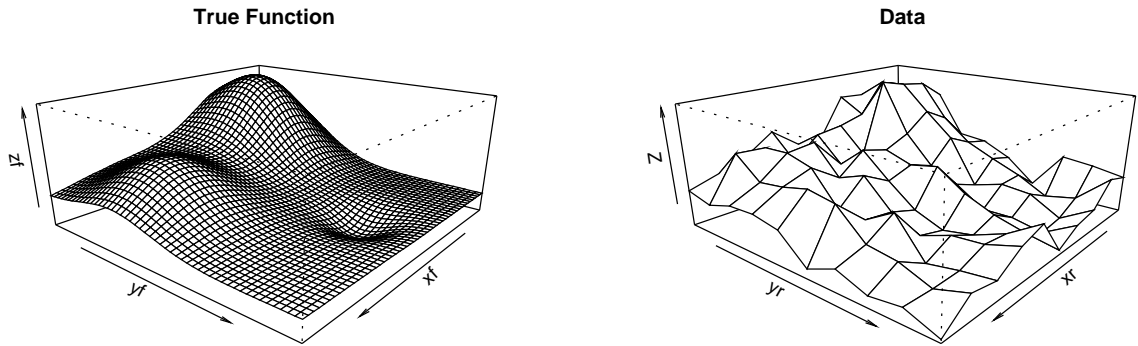}
\caption{True regression function $m(x_1,x_2)$ (\ref{eq:mexican.hat})
on the square $[0,1]\times[0,1]$ used in our numerical examples and a sample of
size 100 with errors and a sample of 100 points}
   \label{fig:chapeau}
\end{center}
\end{figure}

The eigenvalues of the associated 
smoothing matrix lie between zero and one.  In light of proposition
\ref{theorem:converge}, the sequence of bias corrected thin-plate spline
smoothers, starting from a pilot that oversmooths the data, will converge 
to an interpolant of the raw data.  As a result, we anticipate that after some 
suitable number of bias correction steps, the resulting bias corrected
smoother will be a good estimate for the true underlying regression function. 
This behavior is confirmed numerically in the following pedagogical example
of a bivariate regression problem:
Figure \ref{fig:chapeau} graphs Wendelberger's test function
\citep{wendelberger1982}
\begin{eqnarray}
\!\!\!\!\!\!
m(x,y) &=& \frac{3}{4}\exp \left (-\frac{(9x-2)^2 + (9y-2)^2}{4} \right ) 
+\frac{3}{4}\exp \left (-\frac{(9x+1)^2}{49} + \frac{(9y+1)^2}{10} \right )
\nonumber \\
&& \!\!\!\!+\frac{1}{2}\exp \left (-\frac{(9x-7)^2 + (9y-3)^2}{4} \right )
-\frac{1}{5}\exp \left (-(9x-4)^2 - (9y-7)^2 \right ) \label{eq:mexican.hat}
\end{eqnarray}
that is sampled at 100 locations on the regular grid
$\{0.05,0.15,\ldots,0.85,0.95\}^2$.  The disturbances are mean zero
Gaussian with variance producing a signal to noise ratio of five.

\begin{figure}[H]
\begin{center}
\includegraphics{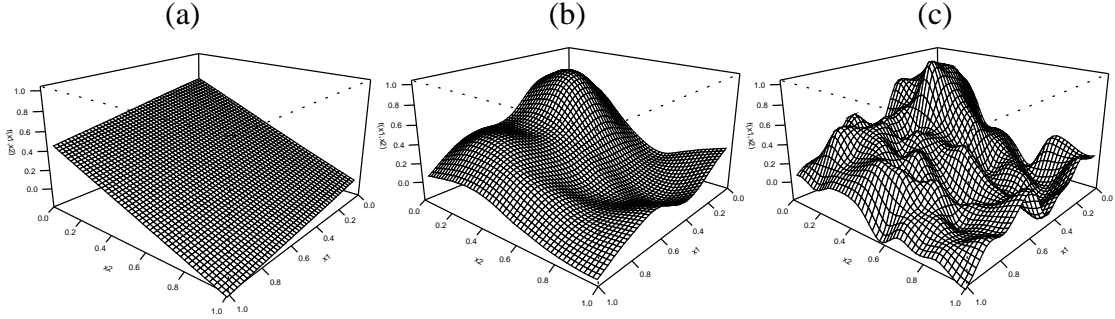}
\caption{TPS regression smoothers from $100$ noisy observations from (\ref{eq:mexican.hat}) 
(see Figure \ref{fig:chapeau}) evaluated on a regular grid on $[0,1]\times[0,1]$.  
Panel (a) shows the pilot smoother, panel (b) graphs the bias corrected smoother after 
500 iterations and panel (c) graphs the smoother after 50000 iterations of the bias 
correction scheme.\label{fig:exemple0}} 
\end{center}
\end{figure}

\vspace*{-.5cm}
Figure \ref{fig:exemple0} shows the evolution of the bias corrected
smoother, starting from a nearly linear pilot smoother in panel (a).
At iteration $k=500$ (or 499 iterative bias reduction steps), the smoother 
shown in panel (b) is visually close to the original regression function.  
Continuing the bias correction scheme will eventually lead to a smoother that
interpolates the raw data. This example shows the importance of suitably 
selecting the number of bias correction iterations.

\subsection{Adaptation to smoothness of the regression function}
Let $\Omega$ be an open bounded subset of $\R^d$ and suppose that  the 
unknown regression function $m$ belongs to the Sobolev space 
$\HH^{(\nu)}(\Omega)=\HH^{(\nu)}$, where $\nu$ is an integer such that 
$\nu>d/2$.    Let $S$ denote the smoothing matrix of a thin-plate spline of order $\nu_0
\leq \nu$ (in practice we will take the smallest possible value 
$\nu_0 = \lfloor d/2 \rfloor +1$)  and fix the smoothing 
parameter $\lambda_0 > 0$ to some reasonably large value.   
Our next theorem states
that there exists a number of iterations $k=k(n)$, depending
on the sample size, for which the resulting estimate $\widehat m_k$ 
achieves the optimal rate of convergence. In light of that theorem,
we expect that an iterative bias corrected smoother, with the number of 
iterations selected by GCV, will achieve the optimal rate of convergence.

\begin{theorem}
\label{adaptative}
Assume that the design $X_i \in \Omega$, $i=1,\ldots,n$
satisfies the following assumption:  Define
\[
h_{max}(n) = \sup_{x \in \Omega} \inf_{i=1,\ldots,n}|x-X_i|,
\mbox{ and } h_{min}(n) = \min_{i\neq j} |X_i-X_j|,
\]
and assume that there exists a constant $B>0$ such that 
\[
\frac{h_{max}(n)}{h_{min}(n)}\leq B \quad \forall n.
\]
Suppose that the true regression function $m \in \HH^{(\nu)}$.\\

If the initial estimator $\hat m_1 = S Y$ is
obtained with $S$ a thin-plate spline of degree
$\nu_0$,  with $\lceil d/2 \rceil  \leq \nu_0<\nu$ and 
a fixed smoothing parameter $\lambda_0 > 0$
not depending on the sample size $n$, 
then there is an optimal number of iterations $k(n)$
such that the resulting smoother $\widehat m_k$ satisfies
\begin{eqnarray*}
\E \left [ \left ( \frac{1}{n}
\sum_{j=1}^n (\hat m_k(X_j) - m(X_j) \right )^2\right ] = O\left(n^{-2\nu/(2\nu+d)}\right),
\end{eqnarray*}
which is the optimal rate of convergence for $m \in  \mathcal{H}^{(\nu)}$.
\end{theorem}

\noindent
While adaptation of the $L_2$-boosting algorithm applied to univariate
smoothing splines  was proven by \citet{buhlmann+2003}, the application
of bias reduction to achieve adaptation to the smoothness of multivariate
regression function has not been previously exploited.  
Rate optimality of the smoother $\hat m_k$ is achieved
by suitable selection of the number of bias correcting iterations, while
the smoothing parameter $\lambda_0$ remains unchanged.   That is, 
the effective size of the neighborhoods the smoother averages 
over remains constant.  Selecting the optimal number of iterations is 
important and we prove that result with GCV criterion using Theorem 
3.2 of \citet{li1987}.

\bigskip
\begin{theorem}
\label{li}
Let $\hat k_{GCV} \in \mathcal{K}_n=\{1,\ldots, \lfloor n^{\gamma}\rfloor \}$, $1 \leq \gamma \leq (2\nu_0)/d$, 
denote the index in the sequence of bias corrected smothers whose
associated smoother minimize the generalized cross-validation criteria.
Suppose that the noise $\varepsilon$ in (\ref{eq:basic.model})
has  finite $4q^{th}$ absolute moment, where $q>\gamma(2\nu/d+1)$, that is, ${\mathbb E}[ |\varepsilon|^{4q} ] < \infty$.
Then as the sample size $n$ grows to infinity,
\begin{eqnarray*} 
\frac{\|\hat m_{\hat k_{GCV}} - m\|^2}{
\inf_{k \in \mathcal{K}_n} \|\hat m_{k} - m\|^2} 
\longrightarrow 1, \quad \hbox{in probability}.
\end{eqnarray*}
\end{theorem}
The moment condition is satisfied for Gaussian or subgaussian errors.

\section{Iterative bias reduction of kernel smoothers}
\label{section:kernel}
The matrix $S$ of thin plate spline is symmetric 
and has eigenvalues in $(0,1]$  
\citep[see for example][]{utreras1988}.  In particular, the first 
$M_0={\nu_0+d-1\choose \nu_0-1}$ eigenvalues are all equal to one, corresponding to the 
parametric component of the smoothing spline.  The sample size $n$
needs to be at least $M_0$, and since from Theorem \ref{adaptative} we 
want $\nu_0 > d/2$, it follows that $M_0$ grows exponentially fast in
the number of covariates $d$.  In particular
the dimension of the parametric component of freedom is 
$5, 28, 165, 1001$ for
$d=4,6,8,10$, respectively, and more generally, $M_0$
grows like $3^{d/2} \times (3/2)^d$ for large $d$.
This feature limits the practical usefulness of TPS smoothers.  For
example, the regression model in Section 5 
for the Boston housing data set 
that has $13$ covariates can not be fit with a TPS
because its sample size $n = 506 < 27500 \approx M_0$.

A possible resolution to this problem is to approximate the
TPS smoother with a kernel smoother, with an appropriate
kernel \citep[see][]{silverman1984,messer1991}.   In this section, 
we discuss kernel based smoothers in general, and we give 
a necessary and sufficient condition on the kernel 
that ensures that the iterative bias correction scheme is well behaved.  
We supplement our theorems with numerical examples of both good and 
bad behavior of our scheme.

\subsection{Kernel type smoothers}
The matrix $S$ of kernel estimators has entries 
$S_{ij} =  K(d_h(X_i,X_j))/\sum_k K(d_h(X_i,X_j))$, where $K(.)$ is 
typically a symmetric function in ${\mathbb R}$ 
(e.g., uniform, Epanechnikov, Gaussian), and $d_h(x,y)$ is a weighted
distance between two vectors 
$x,y \in {\mathbb R}^d$.  The particular choice of the distance
$d(\cdot,\cdot)$ determines the shape of the neighborhood.  For example, 
 the weighted Euclidean norm 
\[
d_h(x,y) = \sqrt{\sum_{j=1}^d \frac{(x_j-y_j)^2}{h_j^2} }, 
\]
where $h=(h_1,\ldots,h_d)$ denotes the bandwidth vector, 
gives rise to elliptic neighborhoods.  

\subsection{Spectrum of kernel smoothers}
While the smoothing matrix $S$ is not symmetric,  it has a real
spectrum.  Write $S= D \mathbb{K}$, where $\mathbb{K}$ 
is symmetric matrix with general element ${\mathbb K}_{ij} = 
K(d_h(X_i,X_j))$ and $D$ is diagonal matrix with elements 
$D_{ii} = 1/\sum_j K(d_h(X_i,X_j))$.  If $q$ is an eigenvector of $S$
associated to the eigenvalue $\lambda$, then
\[
S q = D {\mathbb K} q = D^{1/2} \left ( D^{1/2} {\mathbb K} D^{1/2} \right ) 
D^{-1/2} q = \lambda q,
\]
and hence
\[
\left ( D^{1/2} {\mathbb K} D^{1/2} \right ) \left ( D^{-1/2} q
\right ) = \lambda \left ( D^{-1/2} q \right ).
\]
Hence the symmetric matrix $A=D^{1/2} {\mathbb K} D^{1/2}$ has the
same spectrum as $S$.  Since $S$ is row-stochastic, all its
eigenvalues are bounded by one.  Thus, in light of Theorem 
\ref{theorem:converge}, we seek conditions on the kernel $K$ to
ensure that its spectrum is non-negative.  Necessary and sufficient
conditions on the smoothing kernel $K$ for $S$ to have a non-negative
spectrum are given in the following Theorem.

\begin{theorem} \label{kernel}
If the inverse Fourier-Stieltjes transform of a kernel $K(\cdot)$ 
is  a real positive finite measure,  then the 
spectrum of the Nadaraya-Watson kernel smoother 
lies between zero and one.

Conversely, suppose that $X_1,\ldots,X_n$ are an independent 
$n$-sample from a density $f$ (with respect to Lebesgue measure)
that is bounded away from zero on 
a compact set strictly included in the support of $f$.
If the inverse Fourier-Stieltjes transform of a 
kernel $K(\cdot)$ is not a positive finite measure, then with probability
approaching one as the sample size $n$ grows to infinity, the maximum
of the spectrum of $I-S$ is larger than one.
\end{theorem}

\noindent
{\bf Remark 1:} The assumption that the inverse Fourier-Stieltjes 
transform of a kernel $K(\cdot)$  is  a real positive finite measure
is equivalent to the kernel $K(\cdot)$ being positive-definite function, 
that is, for any finite set of points $x_1,\ldots,x_m$, the matrix
\[
\left ( \begin{array}{ccccc}
K(0) & K(d_h(x_1,x_2)) & K(d_h(x_1,x_3)) & \dots & K(d_h(x_1,x_m)) \\
K(d_h(x_2,x_1)) & K(0) & K(d_h(x_2,x_3)) & \dots & K(d_h(x_2,x_m))\\
\vdots & & & & \vdots \\
K(d_h(x_m,x_1)) & K(d_h(x_m , x_2)) & K(d_h(x_m,x_3)) & \dots & K(0)
\end{array}
\right )
\]
is positive definite.   We refer to \citet{schwartz1993} for a detailed
study of positive definite functions.

\noindent
{\bf Remark 2:} 
\citet{marzio+2008} proved 
the first part of the theorem in the context of
univariate smoothers.  Our
proof of the converse shows that for large enough sample sizes,
most configurations from a random design lead to smoothing matrix
$S$ with negative singular values.   \\

Iterative smoothing of the residuals can be computationally 
burdensome.  To derive an alternative, and computationally 
more efficient representation of the iterative bias corrected 
smoother, observe that
\begin{eqnarray*}
\hat m_k 
&=&[I - D^{1/2}(I - D^{1/2}\mathbb{K}D^{1/2})^kD^{-1/2}]Y\\
&=&D^{1/2}[I-(I - A)^k] D^{-1/2}Y.
\end{eqnarray*}
Writing $A=D^{1/2}\mathbb{K}D^{1/2} = P_{A}\Lambda_AP_A^{t}$,
where $P_A$ is the orthonormal matrix of eigenvectors and 
$\Lambda_A$ diagonal matrix of their associated eigenvalues,
we obtain a computationally efficient representation for the smoother
\begin{eqnarray*}
\hat m_k &=&D^{1/2}P_{A}[I-(I - \Lambda_A)^k]P_A^{t}D^{-1/2}Y.
\end{eqnarray*}
Note that the eigenvalue decomposition of $A$ needs only to
be computed once, and hence leads to a fast implementation 
for calculating the sequence of bias corrected smoothers.

The Gaussian and triangular kernels are positive definite kernels 
(they are the Fourier transform of a finite positive measure, 
\citep{feller1966}).   In light of Theorem \ref{kernel}, the 
iterative bias correction of Nadaraya-Watson kernel 
smoothers with these kernels 
produces a sequence of well behavior smoother. 

\begin{figure}[H]
\begin{center}
\includegraphics{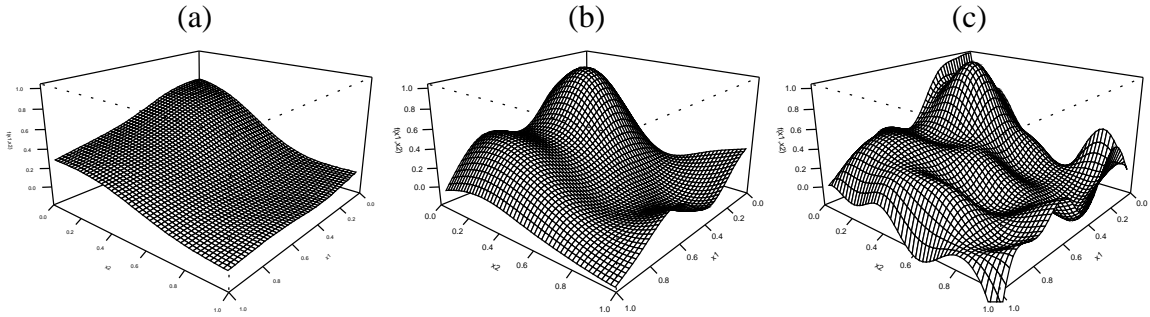}
\caption{Gaussian kernel smoother of $m(x_1,x_2)$ from $n=100$ 
equidistributed points on $[0,1]\times[0,1]$, evaluated on a regular grid
with (a) $k=1$, (b) 50  and (c) 10000 iterations.
\label{fig:exemple1}} 
\end{center}
\end{figure}

The anticipated behavior of iterative bias correction for Gaussian
kernel smoothers is confirmed in our numerical example.  Figure 
\ref{fig:exemple1} shows the progression of the sequence of bias
corrected smoothers starting from a very smooth surface (see panel (a))
that is nearly constant.  Fifty iterations (see panel (b)) produces 
a fit that is visually similar to the original function.  Continued
bias corrections then slowly degrades the fit as the smoother starts
to over-fit the data. 
Continuing the bias correction scheme will eventually lead to a smoother
that interpolates the data.  This example hints at the potential gains
that can be realized by suitably selecting the number of bias correction
steps.

\vspace*{-.5cm}
The uniform and 
the Epanechnikov kernels are not positive definite.  Theorem \ref{kernel}
states that for large enough samples, we expect with high probability 
that $I-S$ has at least one eigenvalue larger than one.  When this
occurs, the sequence of iterative bias corrected smoothers will behave
erratically and eventually diverge.  
Proposition \ref{unif} below strengthens this result by giving
an explicit condition on the configurations of the design points
for which the largest singular value of $I-S$ is always larger 
than one.

\begin{proposition}
\label{unif}
Denote by ${\mathcal N}_i$ the following set: $\{ X_j : K(d_h(X_j,X_i)) > 0 \}$. 

If there exists a set ${\mathcal N}_i$ which contains (at least) two points $X_j,X_k$ different of $X_i$ such that
$d_h(X_i,X_j) < 1$, $d_h(X_i,X_k) < 1$ and $d_h(X_j,X_k) > 1$, then
the smoothing matrix $S$ for the uniform kernel smoother has at least
one negative eigenvalue.

If there exits a set ${\mathcal N}_i$ that contains (at least) two points $X_j,X_k$ ifferent of $X_i$
that satisfy
\[
d_h(X_j,X_k) > \min \{ d_h(X_i,X_j), d_h(X_i,X_k) \},
\]                                  
then the smoothing matrix $S$ for the Epanechnikov kernel smoother has at least
one negative eigenvalue.
\end{proposition}

\begin{figure}[h]
\begin{center}
\includegraphics{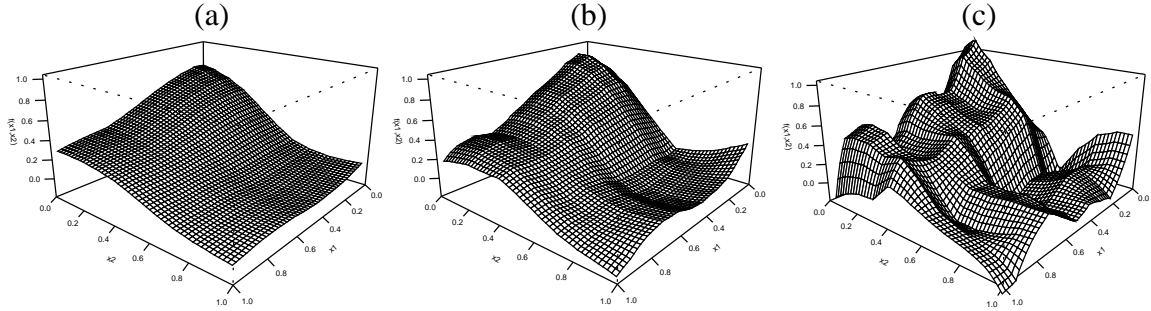}
\caption{Epanechnikov kernel smoother of $m(x_1,x_2)$ from $n=100$ 
 equidistributed points on $[0,1]\times[0,1]$, evaluated on a regular grid
with (a) $k=1$, (b) 5  and (c) 25 iterations.
\label{fig:exemple2}} 
\end{center}
\end{figure}

\noindent
The failure of the iterated bias correction scheme using Epanechnikov kernel 
smoothers is illustrated in the numerical example shown in
Figure \ref{fig:exemple2}.  As for the Gaussian smoother, the initial smoother 
(panel (a)) is nearly constant.  After five iterations (panel (b)) some of the
features of the function become visible.  Continuing the 
bias corrections scheme produces an unstable smoother.  Panel (c) shows that 
after only 25 iterations, the smoother becomes noisy.  Nevertheless, when
comparing panel (a) with panel (b), we see that some improvement is possible
from a few iterations of the bias reduction scheme.

\section{Simulations and a real example}
\label{section:example}
This section presents the results of a modest simulation study
to compare the empirical mean squared error
\begin{eqnarray} \label{eq:mse.sim}
MSE &=& \frac{1}{n} \sum_{i=1}^n(\hat m(X_i) -m(X_i))^2
\end{eqnarray}
of our procedure to its competitors for two functions, in 
dimensions $d=3,5,7$ and sample sizes $n=50,100,200,500,800$,
with a noise to signal ratio of $10\%$. In order to exploit
our theoretical result, the pilot smoother has to oversmooth
otherwise the pilot smoother will have no bias and our
iterative debiasing procedure has no more justification. So
starting with a small $\lambda$ will lead to zero or a small 
number of iterations. Oppositely, starting with a big $\lambda$
will normally lead to a large number of iterations. We decide
in this section to use the values by default in the ibr R-package.
The thin plate spline is govern by a single parameter $\lambda$ that 
weights the contribution of the roughness penalty. For estimating
a $d$-valued regression function, the parametric component is 
$M_0={\nu_0+d-1\choose \nu_0-1}$ and we choose $\lambda$ such that
the initial degree of freedom of the pilot smoother equals equals
$1.5 M_0$. The implementation for the kernel smoother is different
since we could choose a different bandwidth for each explanatory 
variables. We choose one bandwidth for each explanatory variable $X_i$
such as the effective degree of freedom for the one-dimensional smoothing 
matrix related to $X_i$ has a trace equal to 1.1 (more degree than 
a constant but less than a linear model). For such values, the pilot smoothers
always oversmooth.

\begin{figure}[H]
\begin{center}
\includegraphics[width=\textwidth,height=13cm]{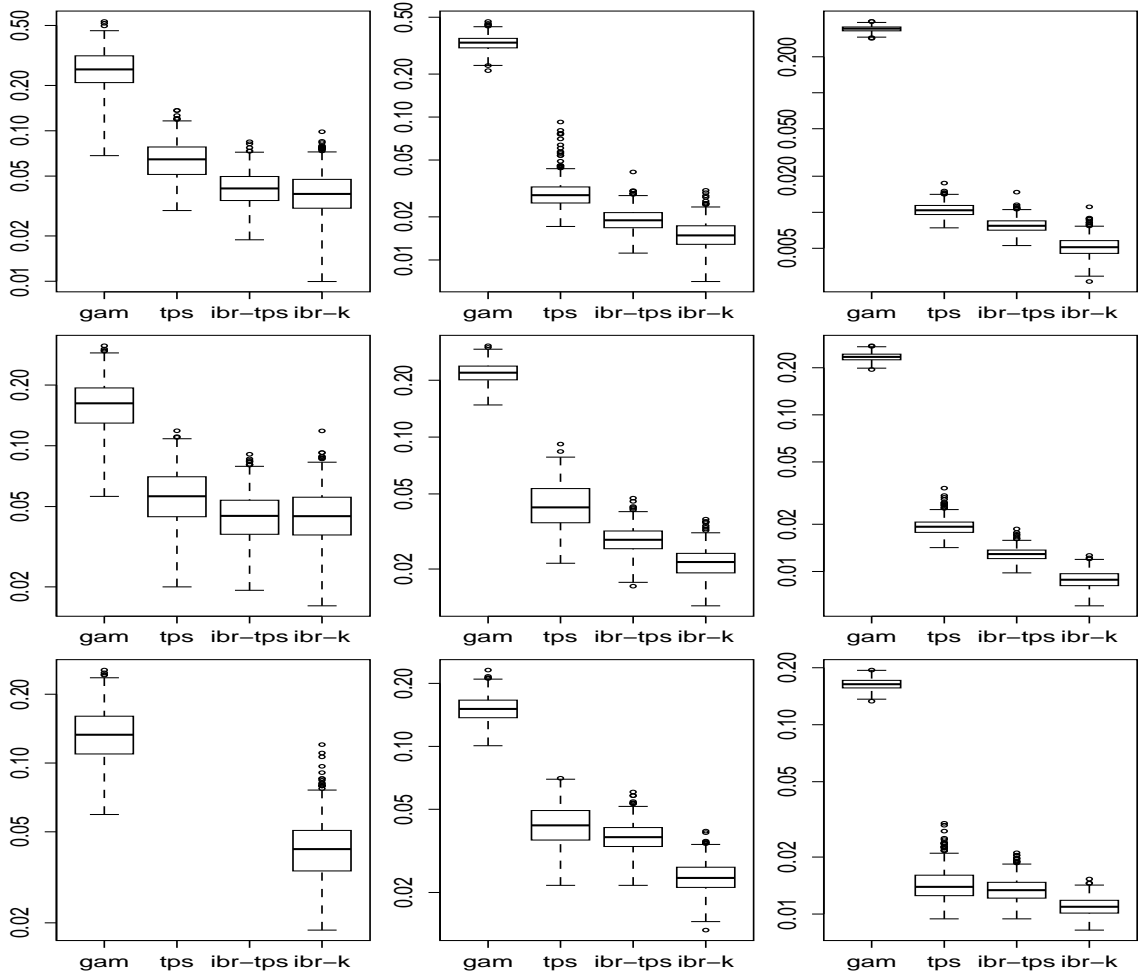}
\caption{Boxplot of Mean Squared Error (MSE) of smoothers for the 
regression functions (from top to bottom) 
of three variables $\sin(2\pi (x_1x_2)^{1/2})+\cos(2\pi
(x_2x_3)^{1/2})$, five variables $\sin(2\pi (x_1x_2x_3)^{1/3})+\cos(2\pi
(x_3x_4x_5)^{1/3})$ and seven variables $\sin(2\pi (x_1x_2x_3x_4)^{1/4})+\cos(2\pi
(x_4x_5x_6x_7)^{1/4})$, and of sample size (from left to right) of
$n=50,200,800$. Each panel shows the boxplot of the 
MSE of a GAM smoother, TPS smoother, IBR with TPS smoother and 
IBR with kernel smoother.   \label{fig:box}} 
\end{center}
\end{figure}

Our simulations was designed to allow us to investigate three aspects:
First, compare the performance of the thin plate spline 
with smoothing parameter selected by GCV
with the IBR smoother using a thin plate spline with a large
smoothing parameter.  We expect that adaptation of our method
will translate into a better performance of our smoother over
the optimal TPS smoother.  Second, to compare the performance
between IBR smoother using either TPS and kernel based smoothers.  
Since kernel 
smoothers do not have a parametric component (which may, or may not,
be needed to fit the data), we believe that kernel smoothers use
more effectively their degree of freedom, which translates into
better performance.  Third, we want to compare the performance 
of fully nonparametric smoothers and additive smoothers.  While 
with additive models we estimate an approximation of the true
regression function, it is generally believed that the approximation
error of an additive model is smaller than the estimation error
of a fully multivariate smoother even for dimensions 
for small sample sizes, e.g. $n=50,100$, and moderate dimensions
of the covariates, e.g. $d=5$.  The results of our study are 
summarized in Table \ref{table:1} and Figure \ref{fig:box}.

Figure \ref{fig:box} shows nine panels each containing 
the boxplots of the MSE  from 500 
simulations, on a logarithmic scale on the $y$-axis. Moving
from top to bottom ranges the regression functions from the
function of three variables $\sin(2\pi (x_1x_2)^{1/2})+\cos(2\pi
(x_2x_3)^{1/2})$, to the function of 
five variables $\sin(2\pi (x_1x_2x_3)^{1/3})+\cos(2\pi
(x_3x_4x_5)^{1/3})$ and to the function of 
seven variables $\sin(2\pi (x_1x_2x_3x_4)^{1/4})+\cos(2\pi
(x_4x_5x_6x_7)^{1/4})$.  All the covariates are i.i.d. uniforms
on the interval $(1,2)$.
Moving from left to right changes
the sample size from $n=50,200,800$.  Within each panel, the 
boxplot of MSE is shown, in the order from left to right,
of additive models using the function {\bf gam}
from the R package {\bf mgcv} , TPS with optimal smoothing 
parameter using the function {\bf
Tps} from the R package {\bf fields}, iterative bias reduction with TPS 
smoother using the function {\bf ibr} from the {\bf ibr} R package  
and iterative bias reduction with kernel smoothers, using again the 
{\bf ibr} function. For reasons explained in Section 4, no TPS smoothers 
can be evaluated for the $d=7$, $n=50$ panel. 

Figure \ref{fig:box} shows that a fully nonparametric smoother
is always preferred to an additive smoother, even for relative
small sample sizes and moderate dimensions.  
\vspace*{-.5cm}
In extensive simulations,
to be reported in a follow-on paper, we observe that this qualitative
conclusion holds over a wide variety of regression functions.
Generally, as expected, the TPS with optimal smoothing 
parameter has a somewhat worse performance than the TPS IBR smoother.  
And finally, the kernel based IBR smoother is 
slightly better than the TPS based IBR smoother, especially in higher 
dimensions.

Table \ref{table:1} gives further insight into the performance
of the various smoothers.  Our table presents the ratio of the
median MSE (in 500 simulation runs) of various smoothers to the
median MSE of the kernel based IBR smoother.  Since
all the entries are larger than one, we conclude that kernel based IBR
consistently outperforms the other smoothing procedures
over the range of sample size, number of covariates and regression functions we
considered in our study.  

\begin{table}[H] \label{table:1}
{\scriptsize
\begin{center}
\begin{tabular}{|c|c|c|c|c|c|}\hline
function & $ n $ &gam & tps & ibr-tps & ibr-k\\ \hline
&50&2.59&1.63&1.39&1\\
&100&4.59&1.89&1.58&1\\
$x_1x_2x_3$&200&8.38&2.14&1.73&1\\
&500&17.9&2.56&2.08&1\\
&800&27.4&2.82&2.39&1\\\hline
&50&6.72&1.70&1.09&1\\
&100&12.0&1.80&1.19&1\\
$\sin(2\pi (x_1x_2)^{1/2})+\cos(2\pi
(x_2x_3)^{1/2})$&200&22.3&1.91&1.27&1\\
&500&46.2&1.99&1.45&1\\
&800&67.3&2.04&1.51&1\\
\hline\hline
&50&2.16&1.60&1.47&1\\
&100&3.83&1.42&1.39&1\\
$x_1x_2x_3x_4x_5$&200&6.64&1.28&1.24&1\\
&500&13.17&1.24&1.22&1\\
&800&19.44&1.26&1.23&1\\
\hline
&50&3.62&1.26&1&1\\
&100&6.32&1.76&1.15&1\\
$\sin(2\pi (x_1x_2x_3)^{1/3})+\cos(2\pi
(x_3x_4x_5)^{1/3})$&200&10.0&1.95&1.31&1\\
&500&18.6&2.06&1.38&1\\
&800&26.5&2.18&1.46&1\\
\hline\hline
&50&2.05&-&-&1\\
&100&3.11&-&-&1\\
$x_1x_2x_3x_4x_5x_6x_7$&200&5.26&3.53&3.17&1\\
&500&9.85&2.46&2.45&1\\
&800&13.8&2.07&2.07&1\\
\hline
&50&3.16&-&-&1\\
&100&4.38&-&-&1\\
$\sin(2\pi (x_1x_2x_3x_4)^{1/4})+\cos(2\pi
(x_4x_5x_6x_7)^{1/4})$&200&6.43&1.78&1.57&1\\
&500&11.1&1.37&1.31&1\\
&800&14.9&1.27&1.22&1\\
\hline\hline
\end{tabular}
\caption{Ratio of median MSE over 500 simulations of a 
smoother and the median MSE over 500 simulations of the kernel 
based IBR smoother. The smoothers, from left to right, are
Generalized Additive Model (GAM), TPS with optimally
selected smoothing parameter (tps), TPS based IBR (ibr-tps) and 
kernel based IBR (ibr-k).}
\end{center}
}
\end{table}
\vspace*{-.5cm}
The improvement over a GAM model  
ranges from 100\% to 6000\%.  This reinforces our conclusions that 
fully nonparametric regressions are practical for moderately large
number of covariates, even for sample sizes as small as $n=50$.

The other notable observation is that the values in the
ibr-tps column are always less than those in the tps column,
showing that consistently, the TPS based IBR smoother has better 
performance than TPS with optimal smoothing parameter.  In our
simulation study, the typical improvement is of 20\%.

\subsection{Boston housing data}
We apply our method on the Boston housing data. This dataset,
created by \citet{harrison+1978} has been extensively to showcase the
performance and behavior of nonparametric multivariate smoothers, 
see for example \citet{breiman+1995} and more recently by \citet{marzio+2008}.
The data contains 13 explanatory variables describing
each of 506 census tracts in the Boston area taken from the
1970 census, together with the median value of owner-occupied  
homes in \$1000's.   The sample size of the data is $n=506$ and the number
of explanatory variables $d=13$. 

We compare our method with the MARS algorithm of \citet{friedman1991} 
as implemented in the  
R package {\bf mda}, with projection pursuit regression (function 
{\bf ppr}), additive models using the backfitting
algorithm of \citet{hastie+1995} as implemented in the R package {\bf mgcv}, 
and additive Boosting \citet{buhlmann+2003} from the R package {\bf mboost}.

The predicted mean squared error is estimated by randomly splitting 30 
times the data into training sets (size $n=350$) and testing sets ($n=156$).
We summarize the results of our analysis in the following table:

\begin{table}[H]
\begin{center}
\caption{Predicted mean Squared Error on test observations for Boston
housing data.\label{tab:mseboston}}
\begin{tabular}{l|l}
Method&Mean Predicted Squared Error\\ \hline\hline
Multivariate regression & 20.09\\ 
$L_2$Boost with component-wise spline & 9.59\\
additive model (backfitted with R) & 11.77\\
Projection pursuit (with R) & 12.64 (4)\\
MARS (with R) &  10.54\\\hline
IBR with GCV stopping rule&\\
and multivariate Gaussian kernel with &\\
{\bf 1.1} initial DDL per variable and {\bf 1230} iterations& 7.35\\
\end{tabular}
\end{center}
\end{table}
\vspace*{-.5cm}
Table \ref{tab:mseboston} again supports our claim  
that the fully multivariate method 
presented in the paper leads to a reduction of more
than 30\% in the prediction mean squared error over
competing state-of-the-art multivariate smoothing 
methods.  A similar comparison for responses on the logarithmic
scale reveals the even larger reduction of 40\% in the 
prediction mean squared error.   Since our fully 
nonparametric regression smoother has substantially smaller
prediction error over additive linear models and low-order
interaction models, we conclude that there exist 
higher order interactions in that data that are significant.

\section{Conclusion}

This paper introduces a fully multivariate regression smoother
for estimating the regression function 
$m(X_1,\ldots,X_d)$ obtained by successive bias
correction from a very smooth (biased) pilot smoother.  We show
that the resulting smoother is adaptive to the underlying smoothness
(see theorems \ref{adaptative} and \ref{li}).
This adaptation to the underlying smoothness 
partially  mitigates the effect from the curse of
dimensionality in many practical examples, and make it practical 
to use fully nonparametric smoother in moderate dimensions, even
for smaller sample sizes.

As in $L_2$ boosting, the proposed iterative bias correction
scheme needs a weak learner
as a base smoother $S$, but all weak learners are not suitable (see
theorem \ref{theorem:converge}).  For instance, 
Epanechnikov kernel smoothers are not interesting (see Theorem
\ref{kernel}).  We further note that one does not
need to keep the same
smoother throughout the iterative bias correcting scheme.  We conjecture
that there are advantages to using weaker smoothers later in the
iterative scheme, and shall investigate this in a forthcoming paper.

Finally, the R package {\bf ibr} available at CRAN implements
the proposed multivariate nonparametric method in R.

\bibliographystyle{abbrvnat} 
\bibliography{./biblio}

\appendix
\section*{Appendix}
\noindent
{\bf Proof of Proposition \ref{theorem:converge}}
\begin{eqnarray*}
\|\hat b_k\|^2 &=& \|-(I-S)^{k-1} SY\|^2\\
&=&  \|(I-S)(I-S)^{k-2} SY\|^2  \leq \|(I-S)\|^2 \|\hat b_{k-1}\|^2\\
& \leq & \|\hat b_{k-1}\|^2,
\end{eqnarray*}
where the last inequality follows from the assumptions on the spectrum of $I-S$.
\medskip \\
\noindent
{\bf Proof of Theorem \ref{adaptative}}
Let $\nu_0<\nu$ and fix the smoothing parameter $\lambda_0$. Define
$S=S_{\nu_0,\lambda_0}$. The eigen decomposition of $S$ \citep{utreras1988}
gives 
\begin{eqnarray*}
\lambda_1=\cdots=\lambda_{M_0}=1 \quad \hbox{and}
\quad \frac{\alpha_j}{1+\lambda_0 j^{2\nu_0/d}} \leq 
\lambda_j \leq \frac{\beta_j}{1+\lambda_0 j^{2\nu_0/d}},
\end{eqnarray*}
where $M_0=\choose^{d+\nu_0-1}_{\nu_0-1}$ and $\alpha_j$ and $\beta_j$ are
two positive constants. We decide to simplify the notation using
\begin{eqnarray*}
\lambda_j \approx \frac{1}{1+\lambda_0 j^{2\nu_0/d}}
\end{eqnarray*}
Let us evaluate the variance of the estimator:
\begin{eqnarray*}
V(\hat m_k,\lambda_0,\nu_0) & \approx & \sigma^2 \frac{M_0}{n}
+ \frac{\sigma^2}{n} \sum_{j=M_0+1}^{n} \left[\left(
1-(1-\frac{1}{1+\lambda_0 j^{2\nu_0/d}})^k\right)\right]^2.
\end{eqnarray*}
Choose $J_n$ in $j=M_0,\ldots,n$, and split the sum in two parts.
Then bound the summand of the first sum by one to get
\begin{eqnarray*}
V(\hat m_k,\lambda_0,\nu_0) &\leq&  \sigma^2 \frac{M_0}{n}
+\sigma^2 \frac{J_n-M_o}{n} 
+\frac{\sigma^2}{n} \sum_{j=J_n+1}^{n} \left[\left(
1-(1-\frac{1}{1+\lambda_0 j^{2\nu_0/d}})^k\right)\right]^2.
\end{eqnarray*}
As the function $1-(1-u)^k \leq k u$ for $u \in [0,1]$, we have
\begin{eqnarray*}
V(\hat m_k,\lambda_0,\nu_0) &\leq& \sigma^2 \frac{J_n}{n} 
+ k^2 \frac{\sigma^2}{n}\sum_{j=J_n+1}^{n} \left(\frac{1}{1+\lambda_0 j^{2\nu_0/d}}\right)^2\\
&\leq& \sigma^2 \frac{J_n}{n} 
+ k^2 \frac{\sigma^2}{n}\sum_{j=J_n+1}^{n} \frac{1}{\lambda_0^2 j^{4\nu_0/d}}.
\end{eqnarray*}
Bounding the sum by the integral and evaluate the latter, one has
\begin{eqnarray*}
V(\hat m_k,\lambda_0,\nu_0) &\leq & \sigma^2 \frac{J_n}{n}
+ k^2 \frac{\sigma^2}{n} \frac{1}{\lambda^2 (4\nu_0/d-1)}
J_n^{-4\nu_0/d+1}.
\end{eqnarray*}
If we want to balance the two terms of the variance, one has to choose
the following number of iterations  $K_n=O(J_n^{2\nu_0/d})$. For such a
choice the variance is of order
\begin{eqnarray*}
V(\hat m_k,\lambda_0,\nu_0) &=& O\left(\frac{J_n}{n}\right).
\end{eqnarray*}
Let us evaluate the squared bias of $\hat m_k$. Recall first the decomposition 
of $S_{\nu_0,\lambda_0} =  P_{\nu_0} \Lambda  P'_{\nu_0}$ and denote by
$\mu_{j,\nu_0} = [P'_{\nu_0}]_j m$ the coordinate of $m$ in the eigen vector 
space of $S_{\nu_0,\lambda_0}$. 
\begin{eqnarray*}
b(\hat m_k,\lambda_0,\nu_0) &=& 
\frac{1}{n}\sum_{j=1}^{n}\left(1-\lambda_j\right)^{2k} \mu_{j,\nu_0}^2\\
&=&\frac{1}{n}\sum_{j=M_0+1}^{j_n}
\left(1-\lambda_j\right)^{2k} \mu_{j,\nu_0}^2+
\frac{1}{n}\sum_{j=j_n+1}^{n}\left(1-\lambda_j\right)^{2k} \mu_{j,\nu_0}^2
\end{eqnarray*}
If $m$ belongs to $ \mathcal{H}^{(\nu)}$ it belongs to $\mathcal{H}^{(\nu_0)}$ 
and we have the following relation by property of $\mathcal{H}^{(\nu_0)}$ 
\begin{eqnarray}
\frac{1}{n} \sum_{j=M_0+1}^n j^{2\nu_0/d}\mu_{j,\nu_0}^2 \leq M < \infty.\label{norm.um}
\end{eqnarray}
Using the fact that $\lambda_j>0$, we have: 
\begin{eqnarray*}
b(\hat m_k,\lambda_0,\nu_0) &\le& \frac{1}{n}\sum_{j=M_0+1}^{j_n}j_n^{-2\nu/d}j_n^{2\nu/d}\mu_{j,\nu_0}^2 + \frac{1}{n}\sum_{j=j_n+1}^{n} j^{-2\nu/d}j^{2\nu/d}\mu_{j,\nu_0}^2\\
b(\hat m_k,\lambda_0,\nu_0) &\le& j_n^{-2\nu/d}\sum_{j=M_0+1}^{j_n}j_n^{2\nu/d}\mu_{j,\nu_0}^2 +  j_n^{-2\nu/d}\frac{1}{n}\sum_{j=j_n+1}^{n}j^{2\nu/d}\mu_{j,\nu_0}^2
\end{eqnarray*}
Using the same type of bound  as in equation~(\ref{norm.um}) we get
\begin{eqnarray*}
b(\hat m_k,\lambda_0,\nu_0) &\le& j_n^{-2\nu/d}M' +  j_n^{-2\nu/d}M''.
\end{eqnarray*}
Thus the bias is of order $O(j_n^{-2\nu/d})$. \\
Balancing the squared bias and the variance lead to the choice
\begin{eqnarray*}
J_n &=& O(n^{1/(1+2\nu/d)})
\end{eqnarray*}
and we obtain the desired optimal rate.\\
\\
\\
{\bf Proof of Theorem \ref{li}} We show that conditions (A.1) to (A.7)
given by \citet{li1987}, in theorem 3.2 are satisfied. To make the
proof self contained, we recall briefly these conditions: 
$(A.1) \ \lim_{n\rightarrow\infty}\sup_{k\in\mathcal{K}_n}
\lambda(S_{k})<\infty, 
\ \ (A.2) \ E(\varepsilon^{4m})<\infty, 
\ \  (A.3) \  \sum_{k\in \mathcal{K}_n}(nR_n(k))^{-m}\rightarrow 0,$\\
where $R_n(k)=\E(\|m_n-\hat m_{k,n}\|^2)/n$, ~ $(A.4): \inf_{k  \in \mathcal{K}_n} n^{-1} \|\hat m_{k} - m\|^2\rightarrow 0,$ in probability.\\
$(A.5)$ for any sequence $\{k_n \in \mathcal{K}_n\}$ such that
$n^{-1}\tr (S_{k_n}S'_{k_n})\rightarrow 0$ we have\\
$\{n^{-1}\tr (S_{k_n})\}^2/\{n^{-1}\tr (S_{k_n}S'_{k_n})\} \rightarrow 0,$ 
$(A.6) \  \sup_{k \in \mathcal{K}_n} n^{-1}\tr (S_{k}) \le \gamma_1 \mbox{ for some }1>\gamma_1>0,$ \\
$(A.7) \  \sup_{k \in \mathcal{K}_n} \{n^{-1}\tr (S_{k})\}^2/\{n^{-1}\tr (S_{k}S'_{k})\} \le \gamma_2 \mbox{ for some }1>\gamma_2>0.$\\
\\
{\bf Conditions (A.1) to (A.4)}\\
The eigen values of $S_k$ (denoted as $\lambda(S_{k})$) are between 0 and
1 $\forall n$, thus the condition (A.1) is fulfilled. To fulfill
condition (A.3) we need to calculate $\sum_{k\in \mathcal{K}_n}
nR_n(k)^{-m}$, where $m$ is an integer to be found,
$m_n=(m(X_1),\ldots,m(X_n))'$ and $\hat m_{k,n}=S_kY$. Using Theorem
\ref{adaptative} we have that for an optimal choice of $k$,
$R_n(k)=\mathcal{O}(n^{d/(2\nu+d)})$. Let us choose $\mathcal{K}_n$ such that its 
cardinal is of order $n^{\gamma}$ ($1\leq \gamma \leq (2\nu_0)/d$), we get the order 
of an upper bound of $\sum_{k\in  \mathcal{K}} nR_n(k)^{-m}$ is $n^{\gamma-\frac{md}{2\nu +d}}$. 
To have (A.3) fulfilled we need that $\gamma-\frac{md}{2\nu +d}<0$,
that is $m>\gamma(2\nu/d+1)$.
Condition (A.4) is satisfied because of Theorem \ref{adaptative}.\\
\\
Conditions (A.5) to (A.7) are related the trace of the matrix $S_k$ and of $S_k^2$.  
Let us recall first some general remarks
\begin{eqnarray*}
\frac{1}{n} \tr (S_k) &=& \frac{1}{n} 
\left(M_0 + 
\sum_{j=M_0+1}^n \left[1 - (1-\lambda_j)^k\right]
\right)
\end{eqnarray*}
where the eigen values $\lambda_j$ are less than 1, bigger than 0 and decreasing. So $\tr (S_k)$ and $\tr (S_k^2)$ are increasing with $k$. By proposition 1, 
$\lim_{k_n \rightarrow \infty}\left(\frac{1}{n}\tr (S_{k_n})\right)=1$ and the same is true
for $\tr (S_{k_n}^2)$ so the choice of the maximal value of $k_n$ is important as it
will be emphasize in the proof. The last general remark is the following
\begin{eqnarray*}
\left(\frac{1}{n} \tr S_k \right)^2 \leq \frac{1}{n} \tr S_k^2 \leq  
\frac{1}{n} \tr S_k \leq 1.
\end{eqnarray*}

Thanks to \citet{utreras1988}, we know that 
\begin{eqnarray*}
\lambda_j &\approx& \frac{1}{1+\lambda_0 j^{\alpha_0}},\quad\alpha_0=\frac{2\nu_0}{d}>1.
\end{eqnarray*}
Let us write 
\begin{eqnarray*}
(1-\lambda_j)^k&=&\left[\frac{\lambda_0 j^{\alpha_0}}{1+\lambda_0 j^{\alpha_0}}\right]^k
= \left(1+\lambda_0^{-1}j^{-\alpha_0}\right)^{-k}
\end{eqnarray*}
So we have
\begin{eqnarray*}
\frac{1}{n } \tr (S_k) &\approx&  \frac{1}{n} M_0 
+ \frac{1}{n} \sum_{j=M_0+1}^n 
\left(1-\left[1+\lambda_0^{-1} j^{-\alpha_0}\right]^{-k}\right) \\
&\approx&  \frac{1}{n} M_0  + \sum_{j=M_0+1}^ng_k(j).
\end{eqnarray*}
We can write 
\begin{eqnarray*}
g_k(j_n) &=& 1-\left[1+\lambda_0^{-1} j_n^{-\alpha_0}\right]^{-k}\\
&=& 1 -\exp{[-k \ln{(1+\lambda_0^{-1} j_n^{-\alpha_0})}]}\\
&=& 1 -\exp{[-k\lambda_0^{-1} j_n^{-\alpha_0}]}.
\end{eqnarray*}
Let us consider the case where $j_n$ tends to infinity. We want to
ensure the following condition: $-k_n j_n^{-\alpha_0}$ tends to zero.
Since $k_n=n^\gamma$ consider $\varepsilon > 0$ such that 
$\varepsilon < \alpha_0 - \gamma$ and assume that 
\begin{eqnarray*}
j_n = O(n^{(\gamma+\varepsilon)/\alpha_0}),
\end{eqnarray*}
then $-k\lambda_0^{-1} j_n^{-\alpha_0}\rightarrow 0$ even when $k$ is at maximum rate 
of order $n^\gamma$. Thus when $n$ grows to infinity, $\forall j\ge j_n$ we have 
the following approximation for $ g_k(j)$:
\begin{eqnarray}
g_k(j)\approx kj^{-\alpha_0}\lambda_0^{-1}\label{eq:approximationg}.
\end{eqnarray}
{\bf Order of an upper bound of $\tr(S_k)/n$}\\
$\forall k \in\mathcal{K}_n$, we have when $n$ grows to infinity:
\begin{eqnarray*}
\frac{1}{n}\tr(S_k)&\approx&\frac{M_0}{n}+\frac{1}{n}
\sum_{j=M_0+1}^{j_n}{g_k(j)}+\frac{1}{n}\sum_{j=j_n+1}^{n}{g_k(j)}\\
&\le&\frac{j_n}{n}+\frac{1}{n}\int_{j_n}^ng_k(j)dj\\
&\le&\frac{j_n}{n}+\frac{kj_n^{1-\alpha_0}}{n(\alpha_0-1)}\lambda_0^{-1}
\end{eqnarray*}
with the last approximation which follows from equation~(\ref{eq:approximationg}). 
Using the fact that the maximum rate for $k_n$ is $O(n^\gamma)$ and that 
$(\gamma,\varepsilon)$ are chosen such that $\gamma+\varepsilon<\alpha_0$ we have 
that an upper bound of $\tr(S_k)/n$ is of order of 
$n^{\frac{\gamma+\varepsilon}{\alpha_0}-1}$.\\
\\
{\bf Order of a lower bound of $\tr(S_k^2)/n$}\\
$\forall k \in\mathcal{K}_n$, we have when $n$ grows to infinity:
\begin{eqnarray*}
\frac{1}{n}\tr(S_k^2)
&\approx&\frac{M_0}{n}+\frac{1}{n}\sum_{j=M_0+1}^{j_n}{g_k^2(j)}+\frac{1}{n}\sum_{j=j_n+1}^{n}{g_k^2(j)}\\
&\ge&\frac{M_0}{n}+\frac{g_k^2(j_n)}{n}(j_n-M_0)+\frac{1}{n}\int_{j_n+1}^{n+1}g^2_k(j)dj\\
&\approx&\frac{M_0}{n}+\frac{j_n-M_0}{n}g_k^2(j)+ 
\frac{k^2\lambda_0^{-2}}{n}\frac{(j_n+1)^{-2\alpha_0+1}}{2\alpha_0-1} - 
\frac{k^2\lambda_0^{-2}}{n}\frac{(n+1)^{-2\alpha_0+1}}{2\alpha_0-1} 
\end{eqnarray*}
with the last approximation which follows from equation~(\ref{eq:approximationg}).  
Using one more time equation~(\ref{eq:approximationg}) we get that 
a lower bound of $\frac{1}{n}\tr(S_k^2)$ if of order of 
$k^2n^{\frac{\gamma+\varepsilon}{\alpha_0}-1-2\varepsilon-2\gamma}$.\\
\\
{\bf Condition (A.5)}\\
Using the previous calculated order, we get that the
order of $(\tr(S_k)/n)^2(\tr(S_k^2)/n)^{-1}$ is of order 
$k^{-2}n^{\frac{\gamma+\varepsilon}{\alpha_0}-1+2\varepsilon+2\gamma}$. 
Recall that $(\gamma,\varepsilon)$ are chosen such that
$\gamma+\varepsilon<\alpha_0$, thus provided that $\varepsilon<1$ the
condition (A.5) of Li's theorem are fulfilled. \\
\\
{\bf Condition (A.6)}\\
The $n$ eigen values are decreasing. Consider now, $j_n=n\zeta $ with
$\zeta$ fixed and less than one. We have that the maximal value of the 
mean of the trace which occurs at $k_n=n^\gamma$ is bounded by
\begin{eqnarray*}
\frac{1}{n}\tr(S_k)&\le&\frac{j_n}{n}+\frac{(n-j_n)}{n}k_nj_n^{-\alpha_0},
\end{eqnarray*}
We can easily show that the last quantity is less than a given value
smaller than 1. The aim of setting $k_n$ equal to $n^\gamma$ is to
ensure that at the border of grid $\mathcal{K}_n$, the smoother is not
identity (ie interpolating). When the smoother is too close to the identity matrix, 
conditions A.6 and A.7 are not fulfilled anymore. Moreover, being very close to 
identity is not interesting in a statistical viewpoint.\\
\\
{\bf Condition (A.7)}\\
We want to analyze
\begin{eqnarray*}
\sup_{k \in \mathcal{K}_n} \frac{\left(n^{-1} \tr (S_{k}) \right)^2}{n^{-1} \tr (S_{k}^2)}
\end{eqnarray*}
Denote by 
\begin{eqnarray*}
\alpha_j(k) &=& \ln(1-(1-\lambda_j)^k),
\end{eqnarray*}
so we have
\begin{eqnarray*}
\frac{1}{n}\tr S_k &=& \frac{1}{n} \sum_j \exp{\alpha_j(k)}\\
\frac{1}{n}\tr S_k^2 &=& \frac{1}{n} \sum_j \exp{2\alpha_j(k)}
\end{eqnarray*}
Let us show that the ratio is an increasing sequence in $k$. Let us
evaluate the sign of the derivative of the ratio. We want to evaluate
the sign of
\begin{eqnarray*}
\sum \! \exp{\alpha_j(k)} 
\sum \! \alpha'_j(k)\exp{\alpha_j(k)} 
\sum \! \exp{2\alpha_j(k)}
\!\! - \!\!
\sum \! \exp{\alpha_j(k)} 
\sum \! \exp{\alpha_j(k)} 
\sum \! \alpha'_j(k) \exp{2 \alpha_j(k)}.
\end{eqnarray*}
Simplifying and dividing by $\sum  \exp{\alpha_j(k)}\sum \exp{2\alpha_j(k)}$
leads to 
\begin{eqnarray*}
\frac{\frac{1}{n} \sum \alpha'_j(k) \exp{\alpha_j(k)}}{\frac{1}{n} \sum \exp{\alpha_j(k)}}
-
\frac{\frac{1}{n} \sum \alpha'_j(k) \exp{2\alpha_j(k)}}{\frac{1}{n} \sum \exp{2\alpha_j(k)}}
\end{eqnarray*}
Rewrite as 
\begin{eqnarray}
\label{der}
\sum \alpha'_j(k) \frac{\beta_j(k)}{\sum \beta_j(k)}
-
\sum \alpha'_j(k) \frac{\beta_j(k)^2}{\sum \beta_j(k)^2}
\end{eqnarray}
The $\{\beta_j(k)\}_j$ is an increasing sequence bounded by one.
It is possible to show that \ref{der} is positive using induction.
The maximum of the quantity under consideration 
grid $\mathcal{K}_n$ is obtained at the border of the
grid. Condition (A.7) can be shown to be fulfilled by
using upper bound of $\tr (S_k)/n$ and lower bound $\tr (S^2_k)/n$.\\
\ \\
\noindent
{\bf Proof of Theorem \ref{kernel}} For notational simplicity, we present
the proof in the univariate case.   Let $X_1,\ldots,X_n$ is an i.i.d.
sample from a density $f$ that is bounded away from zero on a compact
set strictly included in the support of $f$. Consider 
without loss of generality that $f(x) \geq c > 0$ for all $|x| < b$.

We are interested in the sign of the quadratic form $u^\prime Au$ where 
the individual entries $A_{ij}$ of matrix $A$ are equal to
\begin{eqnarray*}
A_{ij} &=&  \frac{K_h(X_i-X_j)}{\sqrt{\sum_l K_h(X_i-X_l)}\sqrt{\sum_lK_h(X_j-X_l)}}.
\end{eqnarray*}
Recall the definition of the scaled kernel $K_h(\cdot) = K(\cdot/h)/h$.
If $v$ is the vector of coordinate 
$v_i=u_i/\sqrt{\sum_l K_h(X_i-X_l)}$ then we have $u^\prime Au=v^\prime \mathbb{K} v$, 
where $\mathbb{K}$ is the matrix with individual entries $K_h(X_i-X_j)$. Thus
any conclusion on the quadratic form
$v^\prime\mathbb{K}v$ carry on to the quadratic form $u^\prime Au$. 
\smallskip
To show the existence of a negative eigenvalue for ${\mathbb K}$,
we seek to construct a vector $U=(U_1(X_1),\ldots,U_n(X_n))$ 
for which we can show that the quadratic form
\[
U^\prime {\mathbb K} U = \sum_{j=1}^n \sum_{k=1}^n U_j(X_j) U_k(X_k) K_h(X_j-X_k)
\]
converges in probability to a negative quantity as the
sample size grows to infinity.   We show the latter by
evaluating the expectation of the quadratic form and 
applying the weak law of large number.
\smallskip
Let $\varphi(x)$ be a real function in $L_2$, define its Fourier 
transform 
\begin{eqnarray*}
\hat \varphi(t) &=& \int e^{-2i\pi t x} \varphi(x) dx
\end{eqnarray*}
and its Fourier inverse by
\begin{eqnarray*}
\hat \varphi_{inv}(t) &=& \int e^{2i\pi t x} \varphi(x) dx.
\end{eqnarray*}
For kernels $K(\cdot)$ that are real symmetric  probability densities, 
we have
\begin{eqnarray*}
\hat K(t) &=& \hat K_{inv}(t).
\end{eqnarray*}
From Bochner's theorem, we know that if the kernel $K(\cdot)$ is not positive
definite, then there exists a bounded symmetric set $A$ of positive 
Lebesgue measure (denoted by $|A|$), such that 
\begin{eqnarray} 
\label{eq:not.pd}
\hat K(t) < 0 \quad \forall t \in A.
\end{eqnarray}
Let $\widehat \varphi(t)\in L_2$ be a real symmetric function
supported on $A$, bounded by $B$ (i.e. $|\widehat \varphi(t)|
\leq B$). Obviously, its inverse Fourier transform 
\[
\varphi(x) = \int_{-\infty}^\infty e^{-2\pi i x t} \widehat \varphi(t) 
dt
\]
is integrable and by virtue of Parseval's identity
\begin{eqnarray*}
\|\varphi\|^2 = \|\widehat \varphi\|^2 \leq B^2 |A| < \infty.
\end{eqnarray*}
Using the following version of Parseval's identity \citep[see][p.620]{feller1966}
\begin{eqnarray*} \label{eq:parceval}
\int_{-\infty}^\infty \int_{-\infty}^\infty \varphi(x) \varphi(y) K(x-y) dx dy 
= \int_{-\infty}^\infty |\widehat \varphi(t)|^2 \hat K(t) dt,
\end{eqnarray*}
which when combined with equation (\ref{eq:not.pd}), leads us to conclude that
\begin{eqnarray*}
\int_{-\infty}^\infty \int_{-\infty}^\infty \varphi(x) \varphi(y) K(x-y) dx dy < 0.
\end{eqnarray*}
Consider the following vector
\[
U = \frac{1}{nh} \left [ 
\begin{array}{c}  
\frac{\varphi(X_1/h)}{f(X_1)} {\mathbb I}(|X_1| < b)\\
\frac{\varphi(X_2/h)}{f(X_2)} {\mathbb I}(|X_2| < b)\\
\vdots\\
\frac{\varphi(X_n/h)}{f(X_n)} {\mathbb I}(|X_n| < b)\\
\end{array}
\right ].
\] 
With this choice, the expected value of the quadratic form is 
\begin{eqnarray*}
\E[Q] & = & \E \left [ \sum_{j,k=1}^n U_j(X_j) U_k(X_k) K_h(X_j-X_k)\right ]\\
&=& \frac{1}{n} \int_{-b}^b \frac{1}{f(s) h^2} \varphi(s/h)^2 K_h(0) ds \\
&&\quad +
\frac{n^2-n}{n^2} \int_{-b}^b \int_{-b}^b \frac{1}{h^2}
\varphi(s/h)\varphi(t/h) K_h(s-t) ds dt\\
&=& I_1 + I_2.
\end{eqnarray*}
We bound the first integral
\begin{eqnarray*}
I_1 & = & \frac{K_h(0)}{nh^2} \int_{-b}^b \frac{\varphi(s/h)^2}{f(s)} ds\\
&\leq& \frac{K_h(0)}{nch} \int_{-b/h}^{b/h} \varphi(u)^2 du\\
&\leq& \frac{B^2 |A| K(0)}{c h^2} n^{-1}.
\end{eqnarray*}
Observe that for any fixed value $h$, the latter can be made arbitrarily small
by choosing $n$ large enough.   We evaluate the second integral by noting 
that
\begin{eqnarray}
I_2 &=& \left ( 1-\frac{1}{n} \right ) h^{-2} \int_{-b}^b \int_{-b}^b \varphi(s/h)
\varphi(t/h) K_h (s-t) ds dt \nonumber \\
&=& \left ( 1-\frac{1}{n} \right ) h^{-2} \int_{-b}^b \int_{-b}^b \varphi(s/h)
\varphi(t/h) \frac{1}{h} K\left ( \frac{s}{h} - \frac{t}{h} \right ) ds dt \nonumber \\
&=& \left ( 1-\frac{1}{n} \right ) h^{-1} \int_{-b/h}^{b/h} \int_{-b/h}^{b/h}
\varphi(u) \varphi(v) K(u-v) du dv. \label{eq:want.negative}
\end{eqnarray}
By virtue of the dominated convergence theorem, the value of the 
last integral converges to $\int_{-\infty}^\infty |\widehat \varphi(t)|^2 
\hat K(t) dt < 0$ as $h$ goes to zero.  Thus for $h$ small enough, 
(\ref{eq:want.negative}) is less than zero,  and it follows that 
we can make ${\mathbb E}[Q] < 0$ by taking $n \geq n_0$, for some large 
$n_0$.   Finally, convergence in probability of the quadratic form to its 
expectation is guaranteed by the weak law of large numbers for $U$-statistics
\citep[see][for example]{grams+1973}.   The conclusion of the theorem follows.\\
\\
\noindent
{\bf Proof of Proposition \ref{unif}}
To handle multivariate case, let each component $h_j$ 
of the vector $h$ be larger than the minimum 
distance between three consecutive points, and 
denote by $d_h(X_{i},X_{j})$ the distance between two vectors. 
For example, if the usual Euclidean distance is 
used, we have
\begin{eqnarray*}
d^2_h(X_{i},X_{j}) &=& \sum_{l=1}^d \left(\frac{X_{il}-X_{jl}}{h_l} \right)^2.
\end{eqnarray*}
The multivariate kernel evaluated at $X_i,X_j$ can be written as
$K(d_h(X_i,X_j))$ where $K$ is univariate.
We are interested in the sign of the quadratic form $u^\prime\mathbb{K}u$ (see proof of theorem \ref{kernel}).  
Recall that if $\mathbb{K}$ is semidefinite positive
then all its principal minor \citep[see][p.398]{horn+1985} 
are nonnegative.   In particular, we can show that $A$ is not semidefinite positive
by producing a $3 \times 3$ principal minor with negative determinant.
To this end, take the principal minor $\mathbb{K}[3]$ obtained by taking the rows 
and columns $(i_1,i_2,i_3)$.    The determinant of $\mathbb{K}[3]$ is
\begin{eqnarray*}
det(\mathbb{K} [3])
&=&K(d_h(0))\left[K(d_h(0))^2 - K(d_h(X_{i_3},X_{i_2}))^2\right]\\
&& \!\!\quad - K(d_h(X_{i_2},X_{i_1}))\times\\
&& \!\!\quad \left[K(d_h(0))K(d_h(X_{i_2},X_{i_1}))-K(d_h(X_{i_3},X_{i_2}))K(d_h(X_{i_3},X_{i_1}))\right]\\
&& \!\!\quad + K(d_h(X_{i_3},X_{i_1}))\times\\
&& \!\!\quad 
\left[K(d_h(X_{i_2},X_{i_1}))K(d_h(X_{i_3},X_{i_2}))-K(d_h(0))K(d_h(X_{i_3},X_{i_1}))\right].
\end{eqnarray*}

Let us evaluate this quantity for the uniform and Epanechnikov kernels.\\
\\
\noindent
{\bf Uniform kernel.}  
Choose 3 points in $\{X_i\}_{i=1}^n$ with index $i_1,i_2,i_3$ such that
\[
d_h(X_{i_1},X_{i_2}) < 1, \quad d_h(X_{i_2},X_{i_3}) < 1, \quad \mbox{and} \quad
d_h(X_{i_1},X_{i_3}) > 1.
\]
With this choice, we readily calculate
\begin{eqnarray*}
det(\mathbb{K} [3])&=&0-K_h(0)\left[K_h(0)^2-0\right]-0<0.
\end{eqnarray*}
Since a principal minor of $\mathbb{K}$ is negative, we conclude that
$\mathbb{K}$ and $A$ are not semidefinite positive.\\
\\
\noindent
{\bf Epanechnikov kernel.}
Choose 3 points $\{X_i\}_{i=1}^n$ with index $i_1,i_2,i_3$, such that 
$d_h(X_{i_1},X_{i_3})>\min(d_h(X_{i_1},X_{i_2});d_h(X_{i_2},X_{i_3}))$ and set 
$d_h(X_{i_1},X_{i_2}) =x \le 1$ and $d_h(X_{i_2},X_{i_3})=y \le 1$. 

Using triangular inequality, we have
\begin{eqnarray*}
det(\mathbb{K} [3])&<& 0.75(0.75^2-K(y)^2)-K(x)(0.75 K(x)-K(y)K(\min(x,y)))\\
&&\ \ 
-K(\min(x,y))K(x)K(y)-0.75K(x+y)^2
\end{eqnarray*}
The right hand side of this equation is a bivariate function of $x$ and $y$.
Numerical evaluations of that function show that small $x$ and $y$ leads to negative value of this function, that is the determinant of $\mathbb{K} [3]$ can be negative.
\begin{figure}[H]
\begin{center}
\includegraphics[height=4cm]{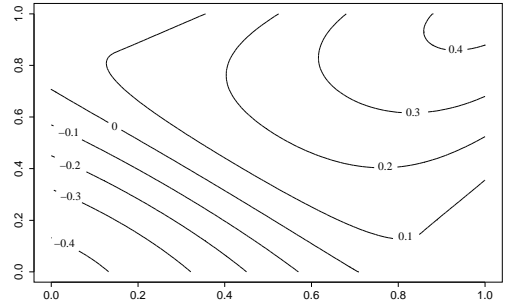}
\caption{Contour of an upper bound of $det(\mathbb{K} [3])$ as a function of $(x,y)$.\label{fig:contour_multi}}
\end{center}
\end{figure}
Thus a principal minor of $\mathbb{K}$ is negative, and as a result,  
$\mathbb{K}$ and $A$ are not semidefinite positive.

\end{document}